\documentclass[aps,pra,reprint,superscriptaddress,showpacs]{revtex4-1}

\usepackage{graphicx}
\usepackage{epstopdf}
\usepackage[colorlinks=true, citecolor=blue, linkcolor=blue, urlcolor=blue]{hyperref}

\begin{document}


\title{Reconfigurable self-sufficient traps for ultracold atoms based on a superconducting square}



\author{M.~Siercke}
\affiliation{Division of Physics and Applied Physics, Nanyang Technological 
University,  21 Nanyang Link, Singapore 637371, Singapore}
\affiliation{Centre for Quantum Technologies, National University of 
Singapore, 3 Science Drive 2, Singapore 117543, Singapore}
\author{K.S.~Chan}
\affiliation{Division of Physics and Applied Physics, Nanyang Technological 
University,  21 Nanyang Link, Singapore 637371, Singapore}
\author{B.~Zhang}
\affiliation{Division of Physics and Applied Physics, Nanyang Technological 
University,  21 Nanyang Link, Singapore 637371, Singapore}
\affiliation{Centre for Quantum Technologies, National University of 
Singapore, 3 Science Drive 2, Singapore 117543, Singapore}
\author{M.~Beian}
\affiliation{Division of Physics and Applied Physics, Nanyang Technological 
University,  21 Nanyang Link, Singapore 637371, Singapore}
\affiliation{Centre for Quantum Technologies, National University of 
Singapore, 3 Science Drive 2, Singapore 117543, Singapore}
\author{M.J.~Lim}
\affiliation{Department of Physics, Rowan University, 201 Mullica Hill 
Road, Glassboro, NJ 08028, USA}

\author{R.~Dumke}
\email{rdumke@ntu.edu.sg}
\affiliation{Division of Physics and Applied Physics, Nanyang Technological 
University,  21 Nanyang Link, Singapore 637371, Singapore}
\affiliation{Centre for Quantum Technologies, National University of 
Singapore, 3 Science Drive 2, Singapore 117543, Singapore}


\date{\today}

\begin{abstract}
We report on the trapping of ultracold atoms in the magnetic field formed entirely by persistent supercurrents induced in a thin film type-II superconducting square. The supercurrents are carried by vortices induced in the 2D structure by applying two magnetic field pulses of varying amplitude perpendicular to its surface. This results in a self-sufficient quadrupole trap that does not require any externally applied fields. We investigate the trapping parameters for different supercurrent distributions. Furthermore, to demonstrate possible applications of these types of supercurrent traps we show how a central quadrupole trap can be split into four traps by the use of a bias field.
\end{abstract}


\maketitle

\section{Indroduction} \label{intro}
Since their first demonstration in 2000 \cite{paper:firstchip}, atom chips have been envisioned to do for matter wave physics what conventional chips have done for electronics. Atoms would be coherently manipulated and steered by microscopic wire patterns on the chip surface. While atom chips have managed to fulfill many of these promises, it has since become apparent that they also bring with them some complications. For chips to manipulate atoms on the micrometer scale they need to confine the atoms close to the wires and the chip surface. This makes the atomic sample susceptible to both current noise in the chip wires and thermal Johnson noise, causing the coherence and trap lifetime of the atoms to be drastically reduced as they are brought closer to the chip \cite{paper:spinflips,paper:johnsonnoise}. Condensates transported in wire guides on atom chips tend to quickly fragment due to wire roughness \cite{paper:ketterlefragmentation, paper:aspectfragmentation, paper:zimmermannfragmentation}, making it challenging to carry out matter-wave interference experiments on-chip. While some of these problems can be solved by adding more complexity to the currents \cite{paper:fragmentationsuppression} or to the chip design \cite{paper:conveyorbelt}, it is clear that they present a significant experimental barrier for the realization of complex on-chip matter-wave manipulation.

In contrast to conventional atom chips, superconducting atom chips promise to solve some of these problems through a significant reduction in the Johnson noise close to the surface \cite{paper:spinfliptheory,paper:FermaniLifetime, paper:zimmermanncoherence}. While they have been shown to improve magnetic trap lifetimes close to the chip surface, they can still suffer from the problem of technical current noise originating from any external power supply controlling the current through the chip wires \cite{paper:haroche, paper:MukaiLifetime}. To produce the long coherence times desirable in many ultracold atom experiments, it therefore becomes necessary to reduce or even eliminate noise from external current sources. One potential solution to this problem is to fabricate the superconducting wires on the chip out of type-II superconducting material. Due to the presence of the mixed state in the phase diagram of these materials, persistent supercurrents can be induced in them by subjecting them to a change in the external magnetic field between the first and second critical fields. These induced supercurrents have been shown to be strong enough to form atomic traps with an additional bias field \cite{paper:dumketrap, paper:dumketraps, paper:shimizutrap}.

\begin{figure}[hb]
 \includegraphics[width=0.9\columnwidth]{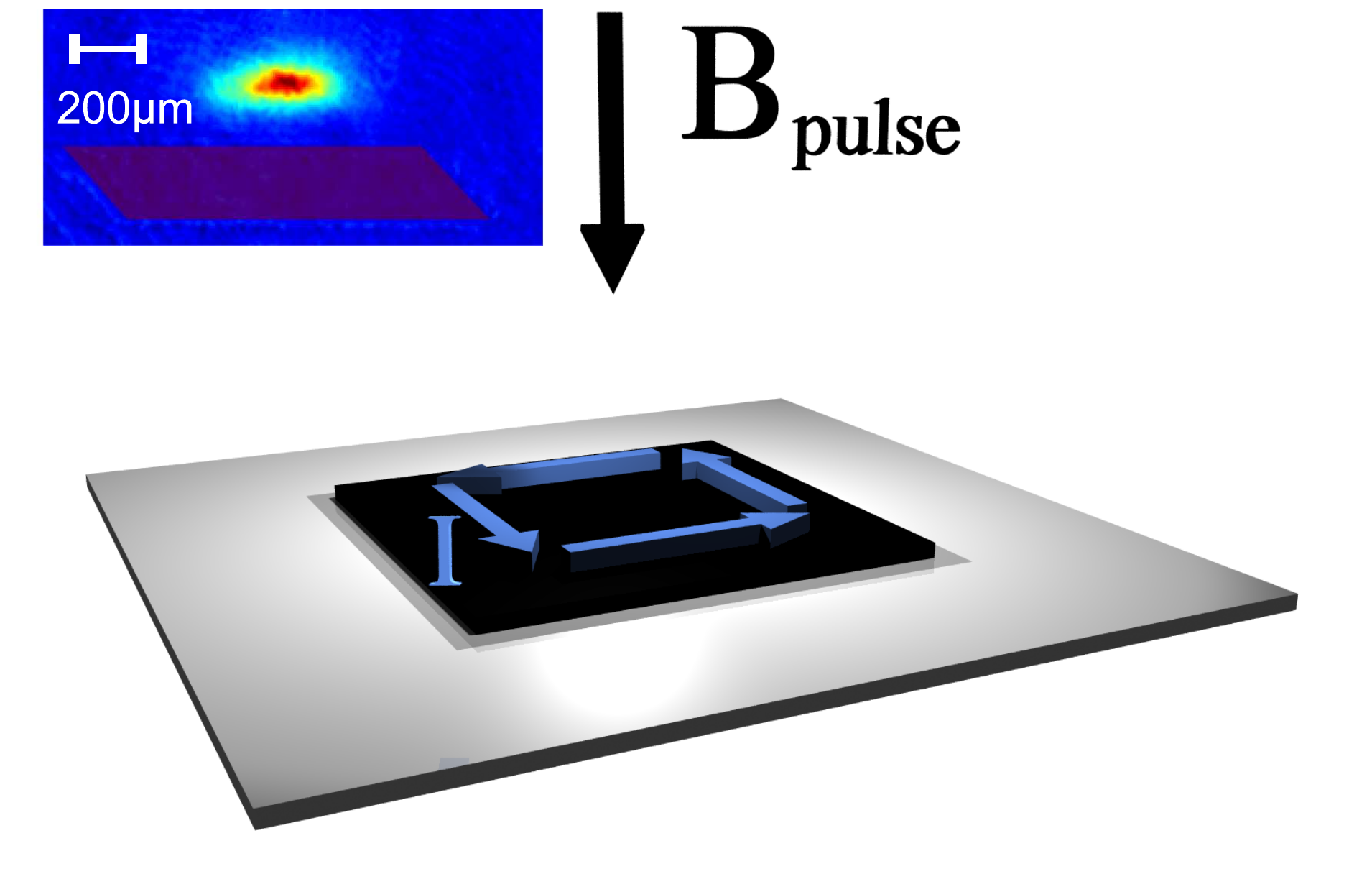}
   \caption{(Color online) Concept of loading supercurrents into a type-II superconducting square. a) Magnetic field pulses are applied perpendicular to the square surface ($B_{pulse}$), inducing vortices in the structure. The vortices carry a supercurrent $I$ giving rise to a magnetic field reflecting the current flow on the surface. Inset: Cloud of $^{87}Rb$ trapped in a quadrupole field produced by the magnetic field from the square and a bias field $B_{bias}=24.5\mathrm{G}$. A single field pulse $B_{pulse}=278\mathrm{G}$ was used to load supercurrents into the square (shaded area). The image is smoothed with a gaussian filter ($\sigma=6.5\mathrm{\mu m}$) to reduce pixel noise.}
\label{fig:chip}
\end{figure}

In this paper we report on the production and loading of a magnetic trap for ultracold atoms formed entirely by persistent supercurrents induced in a superconducting square. We discuss the magnetic field pulse sequence required to form such a self-sufficent trap and give an estimate of its trap depth based on the transfer efficiency from an external quadrupole trap. To demonstrate the versatility of using fields patterned into a type-II superconductor on a chip we show that the fields induced into the same square can act as a 4-way beamsplitter for the atoms, controllable by an external bias field. The methods outlined in this paper may therefore be implemented in designing future low-noise on-chip atom interferometer experiments.

\begin{figure}[!hh]
 \includegraphics[width=0.9\columnwidth]{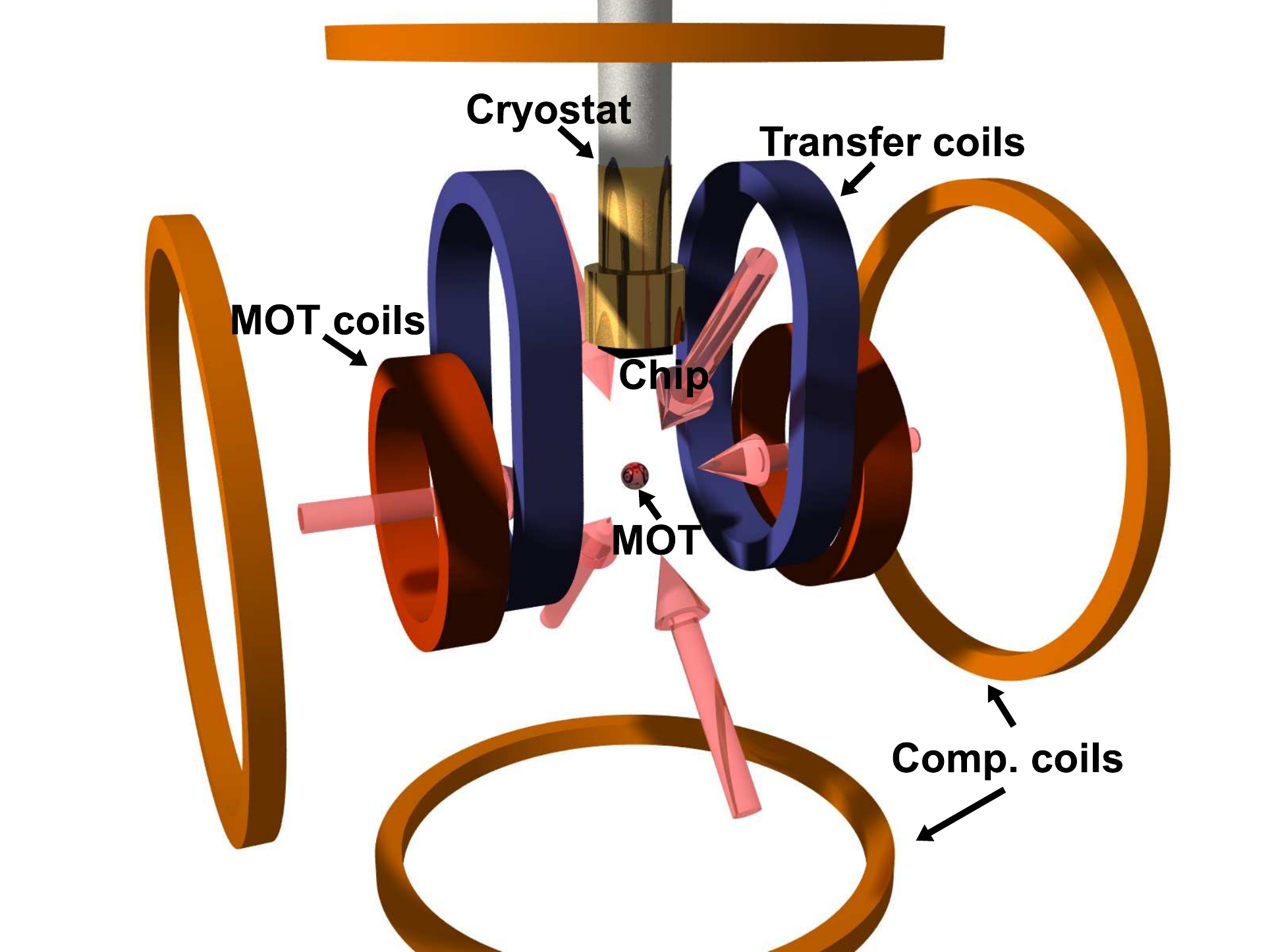}
   \caption{(Color online) The experimental components used to prepare and transport ultracold $^{87}Rb$ atoms to the chip. Atoms are loaded into a magneto-optical trap (MOT), cooled by molasses and trapped in a quadrupole trap after optical pumping. A second set of quadrupole coils transfers the atoms close to the chip where they can be moved in all directions with the use of three helmholtz pairs (compensating coils). For clarity the third set of compensating coils is not shown.}
\label{fig:expt}
\end{figure}

\section{Setup}\label{setup}
The atom chip used in the experiment uses a $800\mathrm{nm}$ thick layer of YBCO as the superconducting material with a critical temperature of $\sim87\mathrm{K}$ patterned onto an Yttria Stabilized Zirconia (YSZ) substrate (figure \ref{fig:chip}). 
The chip is cooled to $83\mathrm{K}$ resulting in a critical current density of $J_c=1\mathrm{MA/cm^2}$. A $1\mathrm{mm}\times1\mathrm{mm}$ square of YBCO is deposited onto the substrate without any leads connecting the structure to an external power source. Once cooled below the superconducting phase transition, vortices are loaded into the square by applying field pulses perpendicular to its surface. The strength of each field pulse $B_{pulse}$ is chosen to be greater than the first critical field of the superconductor but below the second critical field ($B_{c1}<B_{pulse}<B_{c2}$). The pattern of the supercurrent induced in the square by this method is governed by the geometry of the superconductor as well as the sequence of loading pulses and their magnitude and direction \cite{paper:Bo}. The inset of figure \ref{fig:chip} shows atoms trapped in a quadrupole trap generated by an external bias field and the magnetic field of supercurrents induced by a single loading pulse.

Up to $10^8$ $^{87}$Rb atoms are collected in a 6-axis magneto-optical trap (MOT) loaded from  light induced atomic desorption (LIAD) (figure \ref{fig:expt}).
After MOT compression, molasses, and optical pumping, the atoms are trapped in the $|F=2, m_F=2>$ state by the magnetic field of the MOT coils. The $35\mathrm{mm}$ transfer to the chip is accomplished by ramping down the MOT coil current while increasing the current through a second pair of ``transfer coils''. Three sets of helmholtz coil pairs allow for translation of the magnetic trap along all 3 axes in order to center the trap on the superconducting square. The magnetic field pulses used to imprint vortex patterns into the square are generated by applying a current of up to 274A to the MOT coils for a duration of $100\mathrm{ms}$ prior to loading the MOT, corresponding to a field at the chip surface of $600\mathrm{G}$. Increasing the amplitude of a pulse increases the penetration depth of the magnetic field into the superconductor, giving control over the magnitude and radius of the induced current loop. Choosing the polarity of the current applied to the MOT coils controls the direction of the supercurrent. The geometry of the current mimics the shape of the superconductor. The induced currents have previously been measured to be stable over the course of 3 hours \cite{paper:dumketraps}, requiring the pulse loading process to only be carried out once during the course of the experiment. Atoms are brought closer to the chip by increasing the bias field in the direction normal to its surface. Once the gradients generated by the supercurrents in the square become comparable those generated by the transfer coils, the atomic distribution in the trap is considerably modified by the chip field. 
\begin{figure}[!hh]
 \includegraphics[width=0.7\columnwidth]{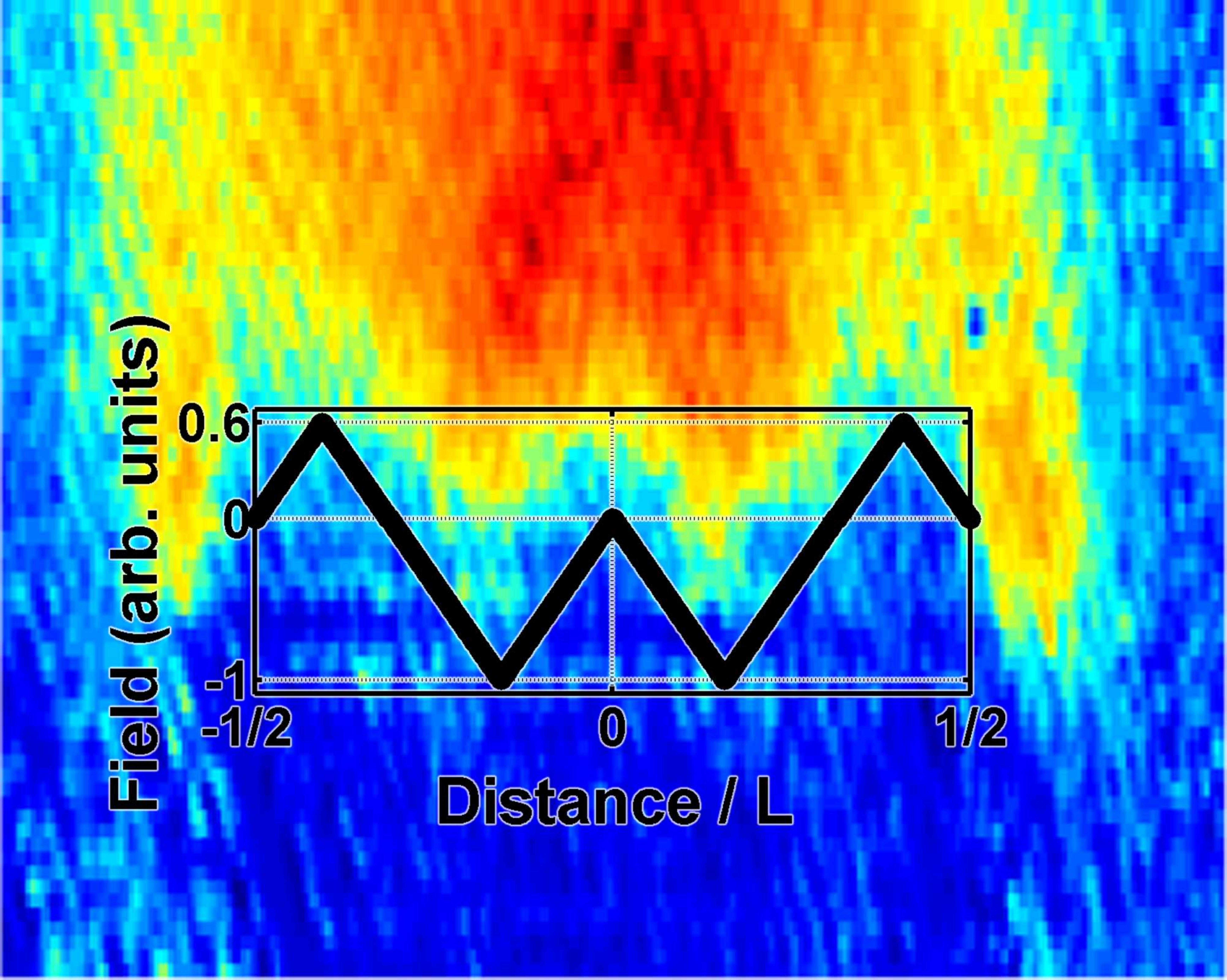}
   \caption{(Color online) Absorption image of ultracold atoms in the vicinity of the superconducting square. The square carries supercurrents loaded by two pulses: $B_{pulse} = 600\mathrm{G}$ and $B_{pulse} = -226\mathrm{G}$. The induced magnetic field along x at the surface of the square  according to Bean's model is shown in black where L is the length of the square. The atoms display a triangular structure closely resembling the magnetic field once they are close to the superconducting surface. The imaging angle is identical to that in the inset of figure \ref{fig:chip}. The image is smoothed with a gaussian filter ($\sigma=6.5\mathrm{\mu m}$) to reduce pixel noise.}
\label{fig:chipatoms}
\end{figure}
In this regime the spatial atomic density distribution reflects the profile of the magnetic field generated by the square, and one can map out the field distribution above the superconductor through absorption imaging of the atoms. Figure \ref{fig:chipatoms} shows the optical density of the atoms in the vicinity of the superconducting square's surface after two magnetic loading pulses: $B_{pulse}=600\mathrm{G}$ followed by $B_{pulse}=-226\mathrm{G}$. The magnetic field structure as predicted by Bean's model \cite{paper:bean} is clearly reflected in the atomic distribution (figure \ref{fig:chipatoms} inset). In this way cold atoms may in the future even be used to image the vortex lattice of the superconductor much as it is currently done with magneto-optical imaging \cite{vorteximaging}.

\section{Results}\label{results}
A Quadrupole trap above the square center can easily be created by loading vortices into the square with a single magnetic field pulse and superimposing an external bias field on the chip field (figure \ref{fig:chip}).
\begin{figure}[!hh]
 \includegraphics[width=0.7\columnwidth]{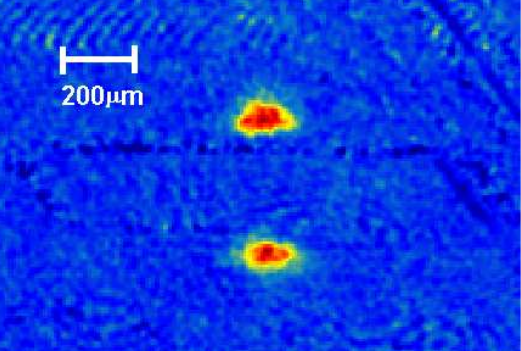}
   \caption{(Color online) Atoms trapped in the magnetic field created by a loading pulse sequence $B_{pulse}=600\mathrm{G}$ followed by $B_{pulse}=-226\mathrm{G}$. The trapping fields are derived entirely from persistent supercurrents on the chip. The bottom cloud is a mirror image of the trapped atoms resulting from the reflective chip surface. The image is smoothed with a gaussian filter ($\sigma=6.5\mathrm{\mu m}$) to reduce pixel noise.}
\label{fig:selfsuff}
\end{figure}
Such a trap still suffers however from field noise arising from current fluctuations in the coils providing the external field. To produce a truly self-sufficient trap the role of the external bias field must be assumed by additional supercurrents induced into the square \cite{paper:Bo}. Since the role of the bias field is to cancel out the field from the square at a certain height $z=z_0$, a second loading field pulse in the direction opposite of the first should induce currents into the square that produce a similar ``bias field''. 

\begin{figure}[h]
 \includegraphics[width=0.9\columnwidth]{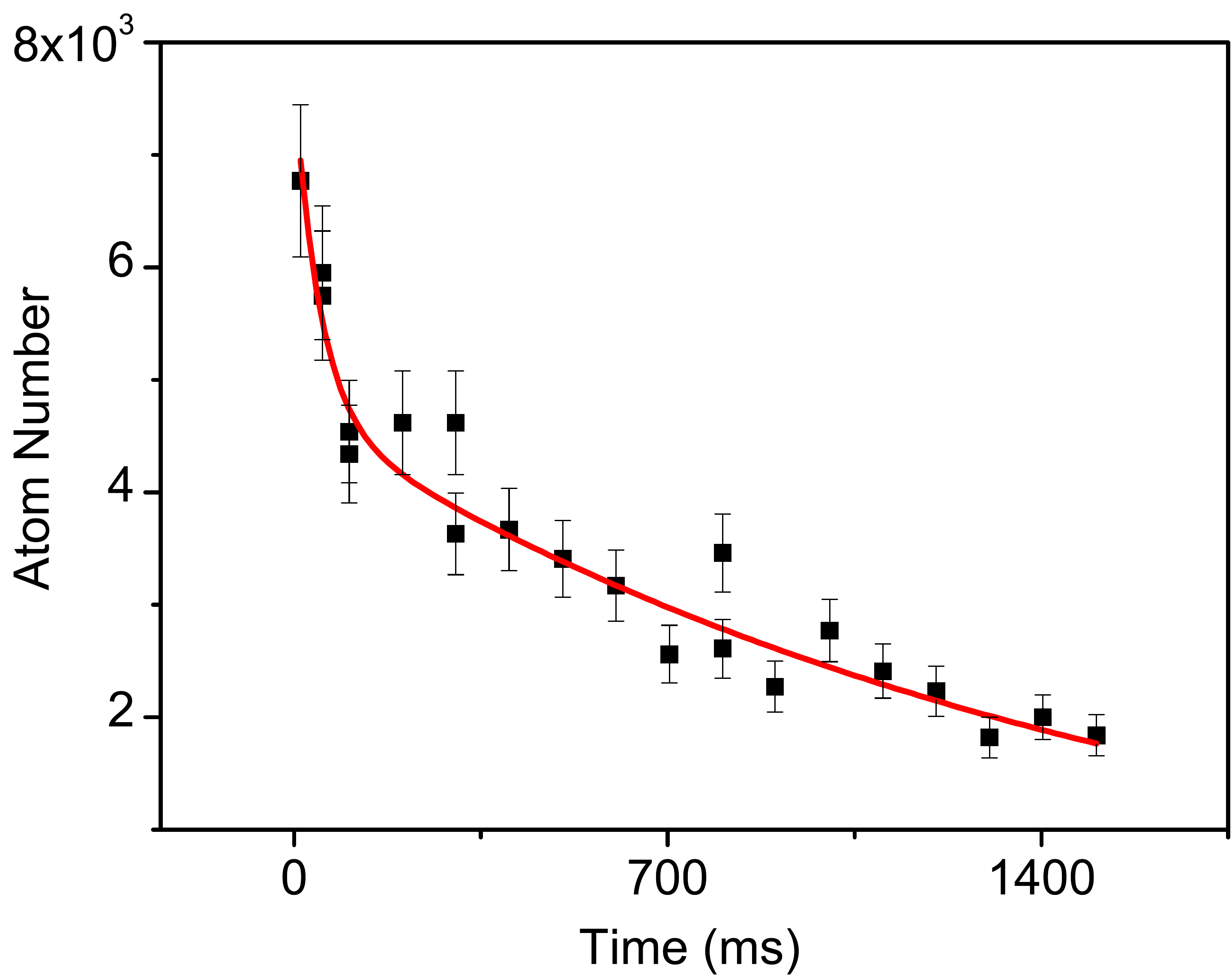}
   \caption{(Color online) Lifetime of the atoms in the self-sufficient trap. The fast initial loss of atoms arises from the fact that the loaded atoms have a temperature higher than the trap depth of $~10$$\mu$K. The decay rate after the initial loss is in agreement with the expected majorana spin-flip rate in the trap \cite{toptrap}. The data is fit to a sum of two exponential decays with no offset.}
\label{fig:lifetime}
\end{figure}

Figure \ref{fig:selfsuff} shows a cloud of $10^3$ ultracold atoms trapped in a self-sufficient quadrupole trap created by such a two-pulse sequence. The distance to the chip surface is $\sim 155\mathrm{\mu m}$. The trap is loaded by moving the center of the magnetic trap produced by the transfer coils to the self-sufficient trap center in $300\mathrm{ms}$, at which point all external fields are switched off in $1\mathrm{ms}$. To determine the precise location of the self-sufficient trap center we make use of the field-mapping technique described in figure \ref{fig:chipatoms} to find the minimum in $|B|$.
The atom number in the trap displays a fast ($50\mathrm{ms}$) initial decay on top of a longer lifetime of $1.5\mathrm{s}$ (figure \ref{fig:lifetime}). 
We attribute the fast initial decay of the atom number to loss of high energy atoms hitting the chip surface. Numerical estimates of the self-sufficient trap depth, as well as evaluations based on the fraction of atoms trapped, give a trap depth of about $10\mathrm{\mu K}$. The temperature of the atoms loaded into the trap on the other hand is $200\mathrm{\mu K}$. The longer, $1.5\mathrm{s}$ lifetime in the trap agrees well with the expected Majorana spin-flip rate of the atoms \cite{toptrap} and could be increased by producing traps on larger squares and by increasing the critical current density by cooling the superconductor further. Following a different route it is also feasible to overcome spin flip losses by dressing the atoms with a radio-frequency field to remove the zero of the magnetic field at the trap center \cite{schmied, spreeuw} or by employing a time-orbiting potential trap \cite{toptrap}. These methods will result in a change of the trapping geometry.

\begin{figure}[!hh]
 \includegraphics[width=\columnwidth]{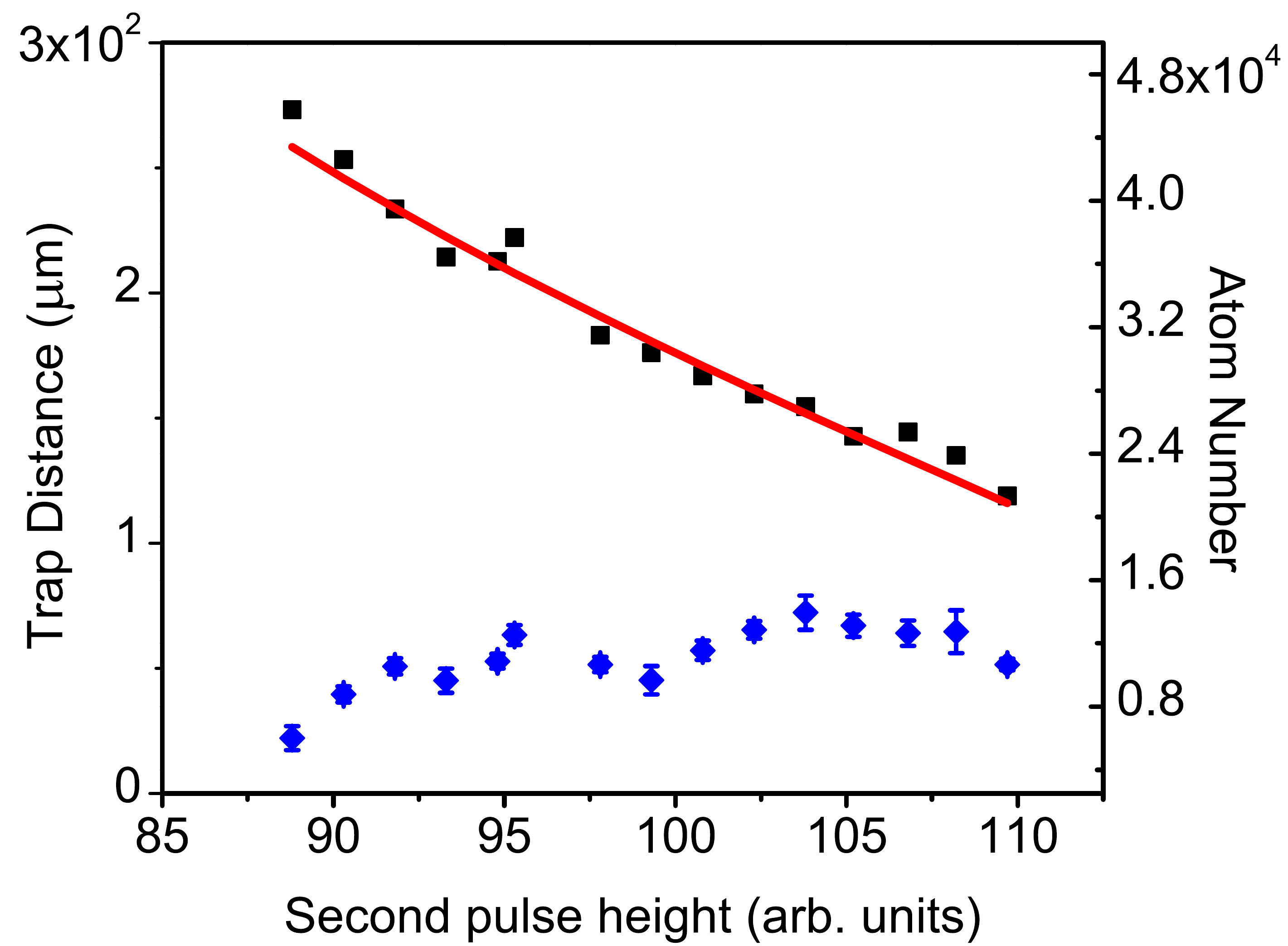}
   \caption{(Color online) Distance of the self-sufficient trap from the superconducting square surface (black squares) and atom number (blue diamonds) versus strength of the second loading pulse. A stronger pulse produces a trap closer to the chip surface. The atom number far away from the chip is reduced due to a shallow trap depth.}
\label{fig:trapdistance}
\end{figure}

The distance of the self-sufficient trap from the chip surface can be varied by changing the relative magnitude of the two vortex-loading pulses. The second, weaker pulse overwrites some of the field of the first pulse and provides the ``bias field'' necessary to produce a field zero above the chip surface. To move the self-sufficient trap further away from the surface it is thus necessary to reduce the magnitude of the second field loading pulse. In doing so the field resulting from the first pulse is made stronger, while the bias field resulting from the second pulse weakens. Figure \ref{fig:trapdistance} shows the distance of the self-sufficient trap from the superconducting surface and the atom number for different strengths of the second loading pulse. To measure the trap distance we make use of the mirrored cloud in figure \ref{fig:selfsuff}.
Since the magnetic field falls off as $1/r^2$ the depth and gradient are reduced for traps far away from the chip. The trap depth and gradient are also reduced in proximity to the chip surface, because the supercurrent cancels the magnetic field at the center of the square. However, the presence of the edges of the superconducting square in the images prevents us from making quantitative measurements of trap distances close to the chip. 

\begin{figure}
 \includegraphics[width=0.7\columnwidth]{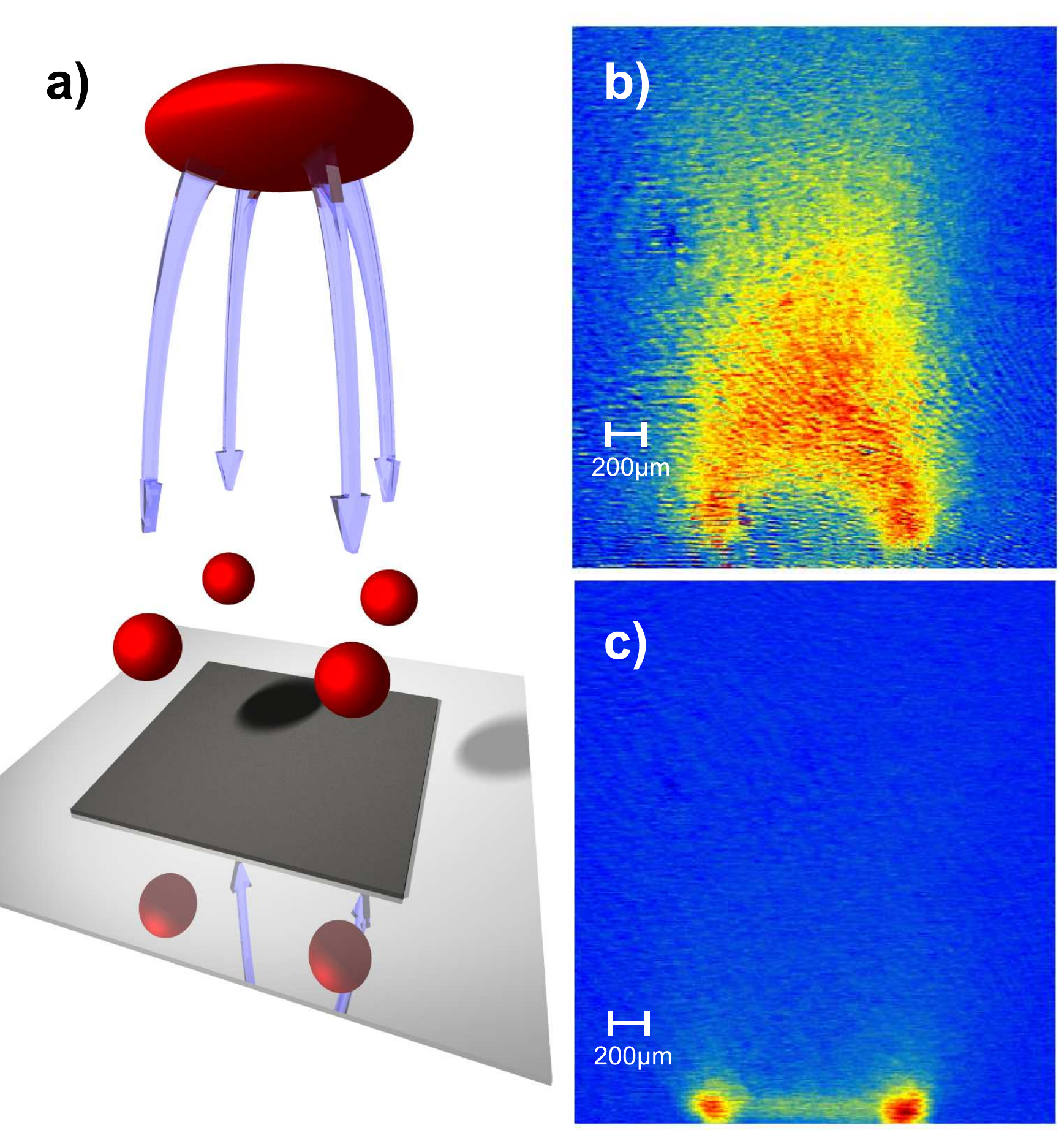}
   \caption{(Color online) a) Magnetic trap geometries formed by two loading pulses $B_{pulse}= 600\mathrm{G}$ and $B_{pulse}= -226\mathrm{G}$ while holding the atoms in an external quadrupole gradient plus an additional bias field $B_{bias}$. b) For low values of $B_{bias}$ there exists one field zero above the square center. c) Increasing the magnitude of $B_{bias}$ brings the atoms closer to the chip surface and splits the initial, central trap into four. Due to the angle of the imaging two of the clouds are hidden behind the other two atom clouds.}
\label{fig:4traps}
\end{figure}

To further demonstrate the flexibility of supercurrents produced by our method we demonstrate controlled splitting of the magnetic traps without the need to structure the underlying material. As numerically simulated in \cite{paper:Bo}, applying a two-pulse sequence such as the one in figure \ref{fig:chipatoms}b followed by an additional bias field, it is possible to produce a ring-shaped trap on top of a superconducting disk without the need to fabricate an on-chip ring structure. In analogy to the case of the disk, supercurrents loaded into a square by this pulse sequence should produce four distinct traps at the four corners of the square. Figure \ref{fig:4traps} shows absorption images of the atoms trapped by this field with an additional bias field $B_{bias}$. 
For low values of $B_{bias}$ a single quadrupole trap is formed above the square center. Increasing the magnitude of the bias field cancels the chip field closer to the surface and the initial, single quadrupole trap is split into four, analogous to a 4-way beamsplitter. Due to the geometry of the imaging system, two of the split clouds are not visible in the figure. By tailoring different loading field sequences or even applying gradients to the superconductor it is possible to create a variety of different trap designs from a single, unpatterned square of type-II superconducting material.

\section{Conclusion}\label{conclusion}
We have shown that, by applying magnetic field pulses perpendicular to the surface of a type-II superconducting square, it is possible to create a fully self-sufficient magnetic trap for ultracold atoms. We estimate the trap depth to be $10\mathrm{\mu K}$ and observe a magnetic trap lifetime due to Majorana spin flips of $1.5$ seconds after an initial loss of atoms due to evaporation by the chip surface. Spin flip losses can potentially be overcome by dressing the atoms with a radio-frequency field to remove the zero of the magnetic field at the trap center \cite{schmied, spreeuw} or by producing a time-orbiting potential trap \cite{toptrap}. However, doing so will result in a change of the trapping geometry. While the trap is loaded from an atomic cloud with a temperature of $200\mathrm{\mu K}$, evaporative cooling prior to loading the self-sufficient trap should result in a higher trapping efficiency. To demonstrate the versatility of writing supercurrent flows into type-II superconductors we show that using the same superconducting square, four distinct quadrupole traps can be formed at the square corners by a two pulse sequence and a bias field. By reducing the bias field the traps can be merged into a single quadrupole trap, resulting in a 4-way splitter for the atoms. The flexible control over the trapping parameters after fabrication, and the low noise inherent in cryogenically cooled chips and supercurrents, give our system distinct advantages over traps generated by normal currents or permanent magnets.


\begin{thebibliography}{22}%
\makeatletter
\providecommand \@ifxundefined [1]{%
 \@ifx{#1\undefined}
}%
\providecommand \@ifnum [1]{%
 \ifnum #1\expandafter \@firstoftwo
 \else \expandafter \@secondoftwo
 \fi
}%
\providecommand \@ifx [1]{%
 \ifx #1\expandafter \@firstoftwo
 \else \expandafter \@secondoftwo
 \fi
}%
\providecommand \natexlab [1]{#1}%
\providecommand \enquote  [1]{``#1''}%
\providecommand \bibnamefont  [1]{#1}%
\providecommand \bibfnamefont [1]{#1}%
\providecommand \citenamefont [1]{#1}%
\providecommand \href@noop [0]{\@secondoftwo}%
\providecommand \href [0]{\begingroup \@sanitize@url \@href}%
\providecommand \@href[1]{\@@startlink{#1}\@@href}%
\providecommand \@@href[1]{\endgroup#1\@@endlink}%
\providecommand \@sanitize@url [0]{\catcode `\\12\catcode `\$12\catcode
  `\&12\catcode `\#12\catcode `\^12\catcode `\_12\catcode `\%12\relax}%
\providecommand \@@startlink[1]{}%
\providecommand \@@endlink[0]{}%
\providecommand \url  [0]{\begingroup\@sanitize@url \@url }%
\providecommand \@url [1]{\endgroup\@href {#1}{\urlprefix }}%
\providecommand \urlprefix  [0]{URL }%
\providecommand \Eprint [0]{\href }%
\providecommand \doibase [0]{http://dx.doi.org/}%
\providecommand \selectlanguage [0]{\@gobble}%
\providecommand \bibinfo  [0]{\@secondoftwo}%
\providecommand \bibfield  [0]{\@secondoftwo}%
\providecommand \translation [1]{[#1]}%
\providecommand \BibitemOpen [0]{}%
\providecommand \bibitemStop [0]{}%
\providecommand \bibitemNoStop [0]{.\EOS\space}%
\providecommand \EOS [0]{\spacefactor3000\relax}%
\providecommand \BibitemShut  [1]{\csname bibitem#1\endcsname}%
\let\auto@bib@innerbib\@empty
\bibitem [{\citenamefont {Zhang}\ \emph {et~al.}(2012)\citenamefont {Zhang},
  \citenamefont {Siercke}, \citenamefont {Chan}, \citenamefont {Beian},
  \citenamefont {Lim},\ and\ \citenamefont {Dumke}}]{paper:Bo}%
  \BibitemOpen
  \bibfield  {author} {\bibinfo {author} {\bibfnamefont {B.}~\bibnamefont
  {Zhang}}, \bibinfo {author} {\bibfnamefont {M.}~\bibnamefont {Siercke}},
  \bibinfo {author} {\bibfnamefont {K.~S.}\ \bibnamefont {Chan}}, \bibinfo
  {author} {\bibfnamefont {M.}~\bibnamefont {Beian}}, \bibinfo {author}
  {\bibfnamefont {M.~J.}\ \bibnamefont {Lim}}, \ and\ \bibinfo {author}
  {\bibfnamefont {R.}~\bibnamefont {Dumke}},\ }\href {\doibase
  10.1103/PhysRevA.85.013404} {\bibfield  {journal} {\bibinfo  {journal}
  {Physical Review A}\ }\textbf {\bibinfo {volume} {85}},\ \bibinfo {pages}
  {013404} (\bibinfo {year} {2012})}\BibitemShut {NoStop}%
\bibitem [{\citenamefont {Dekker}\ \emph {et~al.}(2000)\citenamefont {Dekker},
  \citenamefont {Lee}, \citenamefont {Lorent}, \citenamefont {Thywissen},
  \citenamefont {Smith}, \citenamefont {Drndi\'{c}}, \citenamefont
  {Westervelt},\ and\ \citenamefont {Prentiss}}]{paper:firstchip}%
  \BibitemOpen
  \bibfield  {author} {\bibinfo {author} {\bibfnamefont {N.~H.}\ \bibnamefont
  {Dekker}}, \bibinfo {author} {\bibfnamefont {C.~S.}\ \bibnamefont {Lee}},
  \bibinfo {author} {\bibfnamefont {V.}~\bibnamefont {Lorent}}, \bibinfo
  {author} {\bibfnamefont {J.~H.}\ \bibnamefont {Thywissen}}, \bibinfo {author}
  {\bibfnamefont {S.~P.}\ \bibnamefont {Smith}}, \bibinfo {author}
  {\bibfnamefont {M.}~\bibnamefont {Drndi\'{c}}}, \bibinfo {author}
  {\bibfnamefont {R.~M.}\ \bibnamefont {Westervelt}}, \ and\ \bibinfo {author}
  {\bibfnamefont {M.}~\bibnamefont {Prentiss}},\ }\href {\doibase
  10.1103/PhysRevLett.84.1124} {\bibfield  {journal} {\bibinfo  {journal}
  {Physical Review Letters}\ }\textbf {\bibinfo {volume} {84}},\ \bibinfo
  {pages} {1124} (\bibinfo {year} {2000})}\BibitemShut {NoStop}%
\bibitem [{\citenamefont {Jones}\ \emph {et~al.}(2003)\citenamefont {Jones},
  \citenamefont {Vale}, \citenamefont {Sahagun}, \citenamefont {Hall},\ and\
  \citenamefont {Hinds}}]{paper:spinflips}%
  \BibitemOpen
  \bibfield  {author} {\bibinfo {author} {\bibfnamefont {M.~P.~A.}\
  \bibnamefont {Jones}}, \bibinfo {author} {\bibfnamefont {C.~J.}\ \bibnamefont
  {Vale}}, \bibinfo {author} {\bibfnamefont {D.}~\bibnamefont {Sahagun}},
  \bibinfo {author} {\bibfnamefont {B.~V.}\ \bibnamefont {Hall}}, \ and\
  \bibinfo {author} {\bibfnamefont {E.~A.}\ \bibnamefont {Hinds}},\ }\href
  {\doibase 10.1103/Physrevlett.91.080401} {\bibfield  {journal} {\bibinfo
  {journal} {Physical Review Letters}\ }\textbf {\bibinfo {volume} {91}},\
  \bibinfo {pages} {080401} (\bibinfo {year} {2003})}\BibitemShut {NoStop}%
\bibitem [{\citenamefont {Lin}\ \emph {et~al.}(2004)\citenamefont {Lin},
  \citenamefont {Teper}, \citenamefont {Chin},\ and\ \citenamefont
  {Vuleti\'{c}}}]{paper:johnsonnoise}%
  \BibitemOpen
  \bibfield  {author} {\bibinfo {author} {\bibfnamefont {Y.~J.}\ \bibnamefont
  {Lin}}, \bibinfo {author} {\bibfnamefont {I.}~\bibnamefont {Teper}}, \bibinfo
  {author} {\bibfnamefont {C.}~\bibnamefont {Chin}}, \ and\ \bibinfo {author}
  {\bibfnamefont {V.}~\bibnamefont {Vuleti\'{c}}},\ }\href {\doibase
  10.1103/Physrevlett.92.050404} {\bibfield  {journal} {\bibinfo  {journal}
  {Physical Review Letters}\ }\textbf {\bibinfo {volume} {92}},\ \bibinfo
  {pages} {050404} (\bibinfo {year} {2004})}\BibitemShut {NoStop}%
\bibitem [{\citenamefont {Leanhardt}\ \emph {et~al.}(2002)\citenamefont
  {Leanhardt}, \citenamefont {Chikkatur}, \citenamefont {Kielpinski},
  \citenamefont {Shin}, \citenamefont {Gustavson}, \citenamefont {Ketterle},\
  and\ \citenamefont {Pritchard}}]{paper:ketterlefragmentation}%
  \BibitemOpen
  \bibfield  {author} {\bibinfo {author} {\bibfnamefont {A.~E.}\ \bibnamefont
  {Leanhardt}}, \bibinfo {author} {\bibfnamefont {A.~P.}\ \bibnamefont
  {Chikkatur}}, \bibinfo {author} {\bibfnamefont {D.}~\bibnamefont
  {Kielpinski}}, \bibinfo {author} {\bibfnamefont {Y.}~\bibnamefont {Shin}},
  \bibinfo {author} {\bibfnamefont {T.~L.}\ \bibnamefont {Gustavson}}, \bibinfo
  {author} {\bibfnamefont {W.}~\bibnamefont {Ketterle}}, \ and\ \bibinfo
  {author} {\bibfnamefont {D.~E.}\ \bibnamefont {Pritchard}},\ }\href {\doibase
  10.1103/Physrevlett.89.040401} {\bibfield  {journal} {\bibinfo  {journal}
  {Physical Review Letters}\ }\textbf {\bibinfo {volume} {89}},\ \bibinfo
  {pages} {040401} (\bibinfo {year} {2002})}\BibitemShut {NoStop}%
\bibitem [{\citenamefont {Estev\`{e}}\ \emph {et~al.}(2004)\citenamefont
  {Estev\`{e}}, \citenamefont {Aussibal}, \citenamefont {Schumm}, \citenamefont
  {Figl}, \citenamefont {Mailly}, \citenamefont {Bouchoule}, \citenamefont
  {Westbrook},\ and\ \citenamefont {Aspect}}]{paper:aspectfragmentation}%
  \BibitemOpen
  \bibfield  {author} {\bibinfo {author} {\bibfnamefont {J.}~\bibnamefont
  {Estev\`{e}}}, \bibinfo {author} {\bibfnamefont {C.}~\bibnamefont
  {Aussibal}}, \bibinfo {author} {\bibfnamefont {T.}~\bibnamefont {Schumm}},
  \bibinfo {author} {\bibfnamefont {C.}~\bibnamefont {Figl}}, \bibinfo {author}
  {\bibfnamefont {D.}~\bibnamefont {Mailly}}, \bibinfo {author} {\bibfnamefont
  {I.}~\bibnamefont {Bouchoule}}, \bibinfo {author} {\bibfnamefont {C.~I.}\
  \bibnamefont {Westbrook}}, \ and\ \bibinfo {author} {\bibfnamefont
  {A.}~\bibnamefont {Aspect}},\ }\href {\doibase 10.1103/Physreva.70.043629}
  {\bibfield  {journal} {\bibinfo  {journal} {Physical Review A}\ }\textbf
  {\bibinfo {volume} {70}},\ \bibinfo {pages} {043629} (\bibinfo {year}
  {2004})}\BibitemShut {NoStop}%
\bibitem [{\citenamefont {Fort\'{a}gh}\ \emph {et~al.}(2002)\citenamefont
  {Fort\'{a}gh}, \citenamefont {Ott}, \citenamefont {Kraft}, \citenamefont
  {G\"{u}nther},\ and\ \citenamefont
  {Zimmermann}}]{paper:zimmermannfragmentation}%
  \BibitemOpen
  \bibfield  {author} {\bibinfo {author} {\bibfnamefont {J.}~\bibnamefont
  {Fort\'{a}gh}}, \bibinfo {author} {\bibfnamefont {H.}~\bibnamefont {Ott}},
  \bibinfo {author} {\bibfnamefont {S.}~\bibnamefont {Kraft}}, \bibinfo
  {author} {\bibfnamefont {A.}~\bibnamefont {G\"{u}nther}}, \ and\ \bibinfo
  {author} {\bibfnamefont {C.}~\bibnamefont {Zimmermann}},\ }\href {\doibase
  10.1103/Physreva.66.041604} {\bibfield  {journal} {\bibinfo  {journal}
  {Physical Review A}\ }\textbf {\bibinfo {volume} {66}},\ \bibinfo {pages}
  {041604} (\bibinfo {year} {2002})}\BibitemShut {NoStop}%
\bibitem [{\citenamefont {Trebbia}\ \emph {et~al.}(2007)\citenamefont
  {Trebbia}, \citenamefont {Garrido Alzar}, \citenamefont {Cornelussen}, \citenamefont
  {Westbrook},\ and\ \citenamefont
  {Bouchoule}}]{paper:fragmentationsuppression}%
  \BibitemOpen
  \bibfield  {author} {\bibinfo {author} {\bibfnamefont {J.~B.}\ \bibnamefont
  {Trebbia}}, \bibinfo {author} {\bibfnamefont {C.~L.}\ \bibnamefont
  {Garrido Alzar}}, \bibinfo {author} {\bibfnamefont {R.}~\bibnamefont {Cornelussen}},
  \bibinfo {author} {\bibfnamefont {C.~I.}\ \bibnamefont {Westbrook}}, \ and\
  \bibinfo {author} {\bibfnamefont {I.}~\bibnamefont {Bouchoule}},\ }\href
  {\doibase 10.1103/Physrevlett.98.263201} {\bibfield  {journal} {\bibinfo
  {journal} {Physical Review Letters}\ }\textbf {\bibinfo {volume} {98}},\
  \bibinfo {pages} {263201} (\bibinfo {year} {2007})}\BibitemShut {NoStop}%
\bibitem [{\citenamefont {H\"{a}nsel}\ \emph {et~al.}(2001)\citenamefont
  {H\"{a}nsel}, \citenamefont {Reichel}, \citenamefont {Hommelhoff},\ and\
  \citenamefont {H\"{a}nsch}}]{paper:conveyorbelt}%
  \BibitemOpen
  \bibfield  {author} {\bibinfo {author} {\bibfnamefont {W.}~\bibnamefont
  {H\"{a}nsel}}, \bibinfo {author} {\bibfnamefont {J.}~\bibnamefont {Reichel}},
  \bibinfo {author} {\bibfnamefont {P.}~\bibnamefont {Hommelhoff}}, \ and\
  \bibinfo {author} {\bibfnamefont {T.~W.}\ \bibnamefont {H\"{a}nsch}},\ }\href
  {\doibase 10.1103/PhysRevLett.86.608} {\bibfield  {journal} {\bibinfo
  {journal} {Physical Review Letters}\ }\textbf {\bibinfo {volume} {86}},\
  \bibinfo {pages} {608} (\bibinfo {year} {2001})}\BibitemShut {NoStop}%
\bibitem [{\citenamefont {Hohenester}\ \emph {et~al.}(2007)\citenamefont
  {Hohenester}, \citenamefont {Eiguren}, \citenamefont {Scheel},\ and\
  \citenamefont {Hinds}}]{paper:spinfliptheory}%
  \BibitemOpen
  \bibfield  {author} {\bibinfo {author} {\bibfnamefont {U.}~\bibnamefont
  {Hohenester}}, \bibinfo {author} {\bibfnamefont {A.}~\bibnamefont {Eiguren}},
  \bibinfo {author} {\bibfnamefont {S.}~\bibnamefont {Scheel}}, \ and\ \bibinfo
  {author} {\bibfnamefont {E.~A.}\ \bibnamefont {Hinds}},\ }\href {\doibase
  10.1103/Physreva.76.033618} {\bibfield  {journal} {\bibinfo  {journal}
  {Physical Review A}\ }\textbf {\bibinfo {volume} {76}},\ \bibinfo {pages}
  {033618} (\bibinfo {year} {2007})}\BibitemShut {NoStop}%
\bibitem [{\citenamefont {Fermani}\ \emph {et~al.}(2010)\citenamefont
  {Fermani}, \citenamefont {M\"{u}ller}, \citenamefont {Zhang}, \citenamefont
  {Lim},\ and\ \citenamefont {Dumke}}]{paper:FermaniLifetime}%
  \BibitemOpen
  \bibfield  {author} {\bibinfo {author} {\bibfnamefont {R.}~\bibnamefont
  {Fermani}}, \bibinfo {author} {\bibfnamefont {T.}~\bibnamefont {M\"{u}ller}},
  \bibinfo {author} {\bibfnamefont {B.}~\bibnamefont {Zhang}}, \bibinfo
  {author} {\bibfnamefont {M.~J.}\ \bibnamefont {Lim}}, \ and\ \bibinfo
  {author} {\bibfnamefont {R.}~\bibnamefont {Dumke}},\ }\href {\doibase
  10.1088/0953-4075/43/9/095002} {\bibfield  {journal} {\bibinfo  {journal}
  {Journal of Physics B-Atomic Molecular and Optical Physics}\ }\textbf
  {\bibinfo {volume} {43}},\ \bibinfo {pages} {095002} (\bibinfo {year}
  {2010})}\BibitemShut {NoStop}%
\bibitem [{\citenamefont {Kasch}\ \emph {et~al.}(2010)\citenamefont {Kasch},
  \citenamefont {Hattermann}, \citenamefont {Cano}, \citenamefont {Judd},
  \citenamefont {Scheel}, \citenamefont {Zimmermann}, \citenamefont {Kleiner},
  \citenamefont {Koelle},\ and\ \citenamefont
  {Fort\'{a}gh}}]{paper:zimmermanncoherence}%
  \BibitemOpen
  \bibfield  {author} {\bibinfo {author} {\bibfnamefont {B.}~\bibnamefont
  {Kasch}}, \bibinfo {author} {\bibfnamefont {H.}~\bibnamefont {Hattermann}},
  \bibinfo {author} {\bibfnamefont {D.}~\bibnamefont {Cano}}, \bibinfo {author}
  {\bibfnamefont {T.~E.}\ \bibnamefont {Judd}}, \bibinfo {author}
  {\bibfnamefont {S.}~\bibnamefont {Scheel}}, \bibinfo {author} {\bibfnamefont
  {C.}~\bibnamefont {Zimmermann}}, \bibinfo {author} {\bibfnamefont
  {R.}~\bibnamefont {Kleiner}}, \bibinfo {author} {\bibfnamefont
  {D.}~\bibnamefont {Koelle}}, \ and\ \bibinfo {author} {\bibfnamefont
  {J.}~\bibnamefont {Fort\'{a}gh}},\ }\href {\doibase
  10.1088/1367-2630/12/6/065024} {\bibfield  {journal} {\bibinfo  {journal}
  {New Journal of Physics}\ }\textbf {\bibinfo {volume} {12}},\ \bibinfo
  {pages} {065024} (\bibinfo {year} {2010})}\BibitemShut {NoStop}%
\bibitem [{\citenamefont {Emmert}\ \emph {et~al.}(2009)\citenamefont {Emmert},
  \citenamefont {Lupa\c{s}cu}, \citenamefont {Nogues}, \citenamefont {Brune},
  \citenamefont {Raimond},\ and\ \citenamefont {Haroche}}]{paper:haroche}%
  \BibitemOpen
  \bibfield  {author} {\bibinfo {author} {\bibfnamefont {A.}~\bibnamefont
  {Emmert}}, \bibinfo {author} {\bibfnamefont {A.}~\bibnamefont {Lupa\c{s}cu}},
  \bibinfo {author} {\bibfnamefont {G.}~\bibnamefont {Nogues}}, \bibinfo
  {author} {\bibfnamefont {M.}~\bibnamefont {Brune}}, \bibinfo {author}
  {\bibfnamefont {J.~M.}\ \bibnamefont {Raimond}}, \ and\ \bibinfo {author}
  {\bibfnamefont {S.}~\bibnamefont {Haroche}},\ }\href {\doibase
  10.1140/epjd/e2009-00001-5} {\bibfield  {journal} {\bibinfo  {journal}
  {European Physical Journal D}\ }\textbf {\bibinfo {volume} {51}},\ \bibinfo
  {pages} {173} (\bibinfo {year} {2009})}\BibitemShut {NoStop}%
\bibitem [{\citenamefont {Hufnagel}\ \emph {et~al.}(2009)\citenamefont
  {Hufnagel}, \citenamefont {Mukai},\ and\ \citenamefont
  {Shimizu}}]{paper:MukaiLifetime}%
  \BibitemOpen
  \bibfield  {author} {\bibinfo {author} {\bibfnamefont {C.}~\bibnamefont
  {Hufnagel}}, \bibinfo {author} {\bibfnamefont {T.}~\bibnamefont {Mukai}}, \
  and\ \bibinfo {author} {\bibfnamefont {F.}~\bibnamefont {Shimizu}},\ }\href
  {\doibase 10.1103/Physreva.79.053641} {\bibfield  {journal} {\bibinfo
  {journal} {Physical Review A}\ }\textbf {\bibinfo {volume} {79}},\ \bibinfo
  {pages} {053641} (\bibinfo {year} {2009})}\BibitemShut {NoStop}%
\bibitem [{\citenamefont {M\"{u}ller}\ \emph
  {et~al.}(2010{\natexlab{a}})\citenamefont {M\"{u}ller}, \citenamefont
  {Zhang}, \citenamefont {Fermani}, \citenamefont {Chan}, \citenamefont {Wang},
  \citenamefont {Zhang}, \citenamefont {Lim},\ and\ \citenamefont
  {Dumke}}]{paper:dumketrap}%
  \BibitemOpen
  \bibfield  {author} {\bibinfo {author} {\bibfnamefont {T.}~\bibnamefont
  {M\"{u}ller}}, \bibinfo {author} {\bibfnamefont {B.}~\bibnamefont {Zhang}},
  \bibinfo {author} {\bibfnamefont {R.}~\bibnamefont {Fermani}}, \bibinfo
  {author} {\bibfnamefont {K.~S.}\ \bibnamefont {Chan}}, \bibinfo {author}
  {\bibfnamefont {Z.~W.}\ \bibnamefont {Wang}}, \bibinfo {author}
  {\bibfnamefont {C.~B.}\ \bibnamefont {Zhang}}, \bibinfo {author}
  {\bibfnamefont {M.~J.}\ \bibnamefont {Lim}}, \ and\ \bibinfo {author}
  {\bibfnamefont {R.}~\bibnamefont {Dumke}},\ }\href {\doibase
  10.1088/1367-2630/12/4/043016} {\bibfield  {journal} {\bibinfo  {journal}
  {New Journal of Physics}\ }\textbf {\bibinfo {volume} {12}},\ \bibinfo
  {pages} {043016} (\bibinfo {year} {2010}{\natexlab{a}})}\BibitemShut
  {NoStop}%
\bibitem [{\citenamefont {M\"{u}ller}\ \emph
  {et~al.}(2010{\natexlab{b}})\citenamefont {M\"{u}ller}, \citenamefont
  {Zhang}, \citenamefont {Fermani}, \citenamefont {Chan}, \citenamefont {Lim},\
  and\ \citenamefont {Dumke}}]{paper:dumketraps}%
  \BibitemOpen
  \bibfield  {author} {\bibinfo {author} {\bibfnamefont {T.}~\bibnamefont
  {M\"{u}ller}}, \bibinfo {author} {\bibfnamefont {B.}~\bibnamefont {Zhang}},
  \bibinfo {author} {\bibfnamefont {R.}~\bibnamefont {Fermani}}, \bibinfo
  {author} {\bibfnamefont {K.~S.}\ \bibnamefont {Chan}}, \bibinfo {author}
  {\bibfnamefont {M.~J.}\ \bibnamefont {Lim}}, \ and\ \bibinfo {author}
  {\bibfnamefont {R.}~\bibnamefont {Dumke}},\ }\href {\doibase
  10.1103/PhysRevA.81.053624} {\bibfield  {journal} {\bibinfo  {journal}
  {Physical Review A}\ }\textbf {\bibinfo {volume} {81}},\ \bibinfo {pages}
  {053624} (\bibinfo {year} {2010}{\natexlab{b}})}\BibitemShut {NoStop}%
\bibitem [{\citenamefont {Mukai}\ \emph {et~al.}(2007)\citenamefont {Mukai},
  \citenamefont {Hufnagel}, \citenamefont {Kasper}, \citenamefont {Meno},
  \citenamefont {Tsukada}, \citenamefont {Semba},\ and\ \citenamefont
  {Shimizu}}]{paper:shimizutrap}%
  \BibitemOpen
  \bibfield  {author} {\bibinfo {author} {\bibfnamefont {T.}~\bibnamefont
  {Mukai}}, \bibinfo {author} {\bibfnamefont {C.}~\bibnamefont {Hufnagel}},
  \bibinfo {author} {\bibfnamefont {A.}~\bibnamefont {Kasper}}, \bibinfo
  {author} {\bibfnamefont {T.}~\bibnamefont {Meno}}, \bibinfo {author}
  {\bibfnamefont {A.}~\bibnamefont {Tsukada}}, \bibinfo {author} {\bibfnamefont
  {K.}~\bibnamefont {Semba}}, \ and\ \bibinfo {author} {\bibfnamefont
  {F.}~\bibnamefont {Shimizu}},\ }\href {\doibase
  10.1103/Physrevlett.98.260407} {\bibfield  {journal} {\bibinfo  {journal}
  {Physical Review Letters}\ }\textbf {\bibinfo {volume} {98}},\ \bibinfo
  {pages} {260407} (\bibinfo {year} {2007})}\BibitemShut {NoStop}%
\bibitem [{\citenamefont {Bean}(1964)}]{paper:bean}%
  \BibitemOpen
  \bibfield  {author} {\bibinfo {author} {\bibfnamefont {C.~P.}\ \bibnamefont
  {Bean}},\ }\href {\doibase 10.1103/RevModPhys.36.31} {\bibfield  {journal}
  {\bibinfo  {journal} {Reviews of Modern Physics}\ }\textbf {\bibinfo {volume}
  {36}},\ \bibinfo {pages} {31} (\bibinfo {year} {1964})}\BibitemShut {NoStop}%
\bibitem [{\citenamefont {Goa}\ \emph {et~al.}(2001)\citenamefont {Goa},
  \citenamefont {Hauglin}, \citenamefont {Baziljevich}, \citenamefont
  {Il'yashenko}, \citenamefont {Gammel},\ and\ \citenamefont
  {Johansen}}]{vorteximaging}%
  \BibitemOpen
  \bibfield  {author} {\bibinfo {author} {\bibfnamefont {P.~E.}\ \bibnamefont
  {Goa}}, \bibinfo {author} {\bibfnamefont {H.}~\bibnamefont {Hauglin}},
  \bibinfo {author} {\bibfnamefont {M.}~\bibnamefont {Baziljevich}}, \bibinfo
  {author} {\bibfnamefont {E.}~\bibnamefont {Il'yashenko}}, \bibinfo {author}
  {\bibfnamefont {P.~L.}\ \bibnamefont {Gammel}}, \ and\ \bibinfo {author}
  {\bibfnamefont {T.~H.}\ \bibnamefont {Johansen}},\ }\href {\doibase
  10.1088/0953-2048/14/9/320} {\bibfield  {journal} {\bibinfo  {journal}
  {Superconductor Science \& Technology}\ }\textbf {\bibinfo {volume} {14}},\
  \bibinfo {pages} {729} (\bibinfo {year} {2001})}\BibitemShut {NoStop}%
\bibitem [{\citenamefont {Hofferberth}\ \emph {et~al.}(2006)\citenamefont
  {Hofferberth}, \citenamefont {Lesanovsky}, \citenamefont {Fischer},
  \citenamefont {Verdu},\ and\ \citenamefont {Schmiedmayer}}]{schmied}%
  \BibitemOpen
  \bibfield  {author} {\bibinfo {author} {\bibfnamefont {S.}~\bibnamefont
  {Hofferberth}}, \bibinfo {author} {\bibfnamefont {I.}~\bibnamefont
  {Lesanovsky}}, \bibinfo {author} {\bibfnamefont {B.}~\bibnamefont {Fischer}},
  \bibinfo {author} {\bibfnamefont {J.}~\bibnamefont {Verdu}}, \ and\ \bibinfo
  {author} {\bibfnamefont {J.}~\bibnamefont {Schmiedmayer}},\ }\href {\doibase
  10.1038/nphys420} {\bibfield  {journal} {\bibinfo  {journal} {Nature
  Physics}\ }\textbf {\bibinfo {volume} {2}},\ \bibinfo {pages} {710} (\bibinfo
  {year} {2006})}\BibitemShut {NoStop}%
\bibitem [{\citenamefont {Fernholz}\ \emph {et~al.}(2007)\citenamefont
  {Fernholz}, \citenamefont {Gerritsma}, \citenamefont {Kr\"{u}ger},\ and\
  \citenamefont {Spreeuw}}]{spreeuw}%
  \BibitemOpen
  \bibfield  {author} {\bibinfo {author} {\bibfnamefont {T.}~\bibnamefont
  {Fernholz}}, \bibinfo {author} {\bibfnamefont {R.}~\bibnamefont {Gerritsma}},
  \bibinfo {author} {\bibfnamefont {P.}~\bibnamefont {Kr\"{u}ger}}, \ and\
  \bibinfo {author} {\bibfnamefont {R.~J.~C.}\ \bibnamefont {Spreeuw}},\ }\href
  {\doibase 10.1103/PhysRevA.75.063406} {\bibfield  {journal} {\bibinfo
  {journal} {Physical Review A}\ }\textbf {\bibinfo {volume} {75}},\ \bibinfo
  {pages} {063406} (\bibinfo {year} {2007})}\BibitemShut {NoStop}%
\bibitem [{\citenamefont {Petrich}\ \emph {et~al.}(1995)\citenamefont
  {Petrich}, \citenamefont {Anderson}, \citenamefont {Ensher},\ and\
  \citenamefont {Cornell}}]{toptrap}%
  \BibitemOpen
  \bibfield  {author} {\bibinfo {author} {\bibfnamefont {W.}~\bibnamefont
  {Petrich}}, \bibinfo {author} {\bibfnamefont {M.~H.}\ \bibnamefont
  {Anderson}}, \bibinfo {author} {\bibfnamefont {J.~R.}\ \bibnamefont
  {Ensher}}, \ and\ \bibinfo {author} {\bibfnamefont {E.~A.}\ \bibnamefont
  {Cornell}},\ }\href {\doibase 10.1103/PhysRevLett.74.3352} {\bibfield
  {journal} {\bibinfo  {journal} {Physical Review Letters}\ }\textbf {\bibinfo
  {volume} {74}},\ \bibinfo {pages} {3352} (\bibinfo {year}
  {1995})}\BibitemShut {NoStop}%
\end{thebibliography}
\end{document}